# The X-ray photoelectron and Co K -Edge absorption spectra of $Ba_2CoWO_6$

Dhiraj Jha[1], A. K. Himanshu[2,3,*] , Bijay K Singh[1], Dinesh Shukla [4], Uday Kuma[5] ,Ravi Kumar[6], Kaustava Bhattacharyya[6], A. B. Shinde[6], P. S. R. Krishna[6] , Alo Dutta[7], T. P. Sinha[8], Rajyavardhan Ray[9, 10 #]

[1]Department of Physics, T.M. Bhagalpur University, Bhagalpur-812007, Bihar
[2]Variable Energy Cyclotron Center (VECC), DAE, 1/AF Bidhannagar, Kolkata, India 700064
[3]Homi Bhabha National Institute, Mumbai, India 400094
[4]UGC-DAE-CSR, Indore-Madhya Pradesh-452017, India
[5]Department of Physics, NIT Jamshedpur, Jamshedpur, India
[6]Bhabha Atomic Research Center (BARC), Trombay, Mumbai, India 400085
[7]S N Bose Institute, Kolkata, India -700106
[8]Bose Institute, 93/1, APC Road, Kolkata, India 700009
[9]IFW Dresden, Helmholtzstr. 20, 01069 Dresden, Germany
[10]Dresden Center for Computational Materials Science (DCMS), TU Dresden, 01062 Dresden, Germany

**Corresponding authors**. [#]r.ray@ifw-dresden.de (RR), [*]akhimanshu@gmail.com(AKH).

**Abstract**

The structural and electronic properties of cubic double perovskite $BaCoWO_6$ have been investigated using the X-ray photoelectron spectroscopy (XPS) and X-ray absorption spectroscopy (XAS) . While Extended X-ray Absorption Fine structure (EXAFS) studies provide structural information regarding the local environment of Co in $Ba_2CoWO_6$, the combined XPS and Co-K edge XAS spectra of $Ba_2CoWO_6$ (using Co-foil as a reference) elucidate the electronic properties. We find that the oxidation states for Co, Ba and W can be unambiguously assigned to +2, +2 and +6, respectively.

## Introduction

Double perovskites systems have attracted a lot of attention over the years as they exhibit a wide variety of physical phenomena including colossal magnetoresistance and half-metallicity [1,2]. Therefore, they are considered promising candidates for magneto-electronic technology and spintronic applications, such as magnetic sensors, read-head, and solid oxide fuel cells. It is expected that antiferromagnetic perovskite metals should, in general, be good candidates for exhibiting positive colossal magnetoresistance, at least in the strong coupling limit . The cobalt-tungsten double perovskite systems, represented with the general formula $A_2CoWO_6$ (A = Ca, Sr, Ba), are among the most widely studied half-metallic double perovskites [3 – 10].

$Ba_2CoWO_6$ (BCoW) is known to crystallize into a cubic Fm-3m structure [3,4]. The ground state is an antiferromagnet with $T_N$ = 19 K [3,4]. Spin-polarized Density Functional Theory (DFT) based calculations with GGA and GGA+$U$ suggest that the ground state is a half-metal, where the half-metallicity originates from the fact that the Co-d shell is more than half-filled, leading to a gapped up-channel and large density of states at the Fermi energy in the down-spin channel [5,6]. Within this description, Co atoms are in d7 configuration with a spin moment of approximately 3 $\mu_B$. However, it is observed that the orbital moment in this [7] and related compounds $Sr_2CoWO_6$ [11] is unquenched, as suggested by observed magnetic moment of 4.90 < $\mu_B$ <5.20 for Co2+ configuration. For unquenched orbital moments in a cubic environment, local distortions are expected to play a crucial role.

In this *brief report*, we, therefore, focus on understanding the electronic properties and local environment of Co atoms in BCoW. We present our findings on the structural and electronic properties of BCoW as revealed through X-ray photoelectron and X-ray absorption spectroscopy studies.

**Methods**

BCoW were synthesized using the solid state reactions using stoichiometric mixtures of BaCO3, CoO and WO3 heated at 1200 °C for eight hours. The samples were yellowish dark green in color. Room temperature X-ray diffraction was carried out to confirm the structure. The X-ray diffraction data was analyzed using Rietveld analysis software FULLPROF. The structure is found to be cubic Fm-3m (No. 225) with lattice constant a=8.210Å. The Wyckoff positions of Ba, Co, W, and O atoms are found to be *8c*, *4a*, *4b* and *24e*, respectively.

The XAS measurements have been carried out at the Energy-Scanning EXAFS beamline (BL-9) at the Indus-2 Synchrotron Source (2.5 GeV, 100 mA) at Raja Ramanna Centre for Advanced Technology (RRCAT), Indore, India [12,13]. This beamline operates in the energy range of 4 KeV to 25 KeV. The beamline optics consists of a Rh/Pt coated collimating meridional cylindrical mirror and the collimated beam reflected by the mirror is monochromatized by a Si(111) (2d=6.2709 Å) based double crystal monochromator (DCM). The second crystal of DCM is a sagittal cylinder used for horizontal focusing while a Rh/Pt coated bendable post mirror facing down is used for vertical focusing of the beam at the sample position. Rejection of the higher harmonics content in the X-ray beam is performed by detuning the second crystal of DCM. In the present case, XAS measurements have been performed in both transmission mode and fluorescent mode.

For the transmission measurement, three ionization chambers (300 mm length each) have been used for data collection, one ionization chamber for measuring incident flux ($I_0$), second one for measuring transmitted flux ($I_t$) and the third ionization chamber for measuring XAS spectrum of a reference metal foil for energy calibration. Appropriate gas pressure and gas mixtures have been chosen to achieve 10-20% absorption in first ionization chamber and 70-90% absorption in second ionization chamber to improve the signal to noise ratio. The absorption coefficient $\mu$ is obtained using the relation:

$$I_T = I_0 e^{-\mu x} \quad (1)$$

where, $x$ is the thickness of the absorber.

For measurements in the fluorescence mode, the sample is placed at 45° to the incident X-ray beam, and a fluorescence detector is placed at right angle to the incident X-ray beam to collect the signal. One ionization chamber detector is placed prior to the sample to measure the incident flux ($I_0$) and florescence detector measures the fluorescence intensity ($I_f$). In this case the X-ray absorption co-efficient of the sample is determined by $\mu = I_f / I_0$, and the spectrum was obtained as a function of energy by scanning the monochromator over the specified range.

To obtain the qualitative information about the local structure, oscillations in the absorption spectra $\mu(E)$ has been converted to absorption function $\chi(E)$ defined as follows:

$$\chi(E) = \frac{\mu(E) - \mu_0(E)}{\Delta\mu_0(E_0)} \quad (2)$$

where, $E_0$ is absorption edge energy, $\mu_0(E_0)$ is the bare atom background and $\Delta\mu_0(E_0)$ is the step in $\mu(E)$ value at the absorption edge. The energy dependent absorption coefficient $\chi(E)$ has been converted to the wave number dependent absorption coefficient $\chi(k)$ using the relation,

$$K = \sqrt{\frac{2m(E - E_0)}{\hbar^2}} \quad (3)$$

where, m is the electron mass. $\chi(k)$ is weighted by $k^2$ to amplify the oscillation at high $k$ and the $\chi(k)k^2$ functions are Fourier transformed in $R$ space to generate the $\chi(R)$ versus $R$ spectra in terms of the real distances from the center of the absorbing atom. The set of EXAFS data analysis programme available within Demeter[3] software package have been used for EXAFS data analysis. This includes background reduction and Fourier transform to derive the $\chi(R)$ versus $R$ spectra from the absorption spectra (using ATHENA [14] software), generation of the theoretical EXAFS spectra starting from an assumed crystallographic structure and finally fitting of experimental data with the theoretical spectra using ARTEMIS software [14].

## Results and discussion

### A. X-ray photoelectron spectroscopy (XPS):

Figure 1(a) shows the XPS spectrum of the material in the energy window of 0–900eV. The profiles of the XPS spectra are identified and indexed. The 2p core level spectrum of Co is shown in Fig. 1(b). Due to the spin-orbit effect the 2p spectrum is split into $2p_{1/2}$ and $2p_{3/2}$ peaks, positioned at 798.4 and 782.7 eV, respectively. The binding energy difference between these two spin-orbit peaks is ~15.7 eV. The presence of strong satellite of $2p_{1/2}$ peak at ~ 804.7 eV suggests the presence of $Co^{2+}$ state in this material. The FWHM of $2p_{1/2}$ is found to be ~ 4 eV. The broadening of $2p_{1/2}$ peak is due to the multiple splitting as observed for CoO [15]. In octahedral co-ordination $Co^{2+}$ ions have the $t_{2g}^5e_g^{\,2}$ configuration and S = 3/2. As a result the d-d and O(2p) to Co(3d) charge transfer excitations may arise that enhance the number of final states, thus resulting in the multiplet structure of the Co 2p doublet and satellite lines. The line broadening and the high value of doublet separation (15.7 eV) are responsible for the weak resolution of the multiplet for this material. The chemical shift of Co-2p$_{3/2}$ state is +4.7 with respect to its literature value [16].

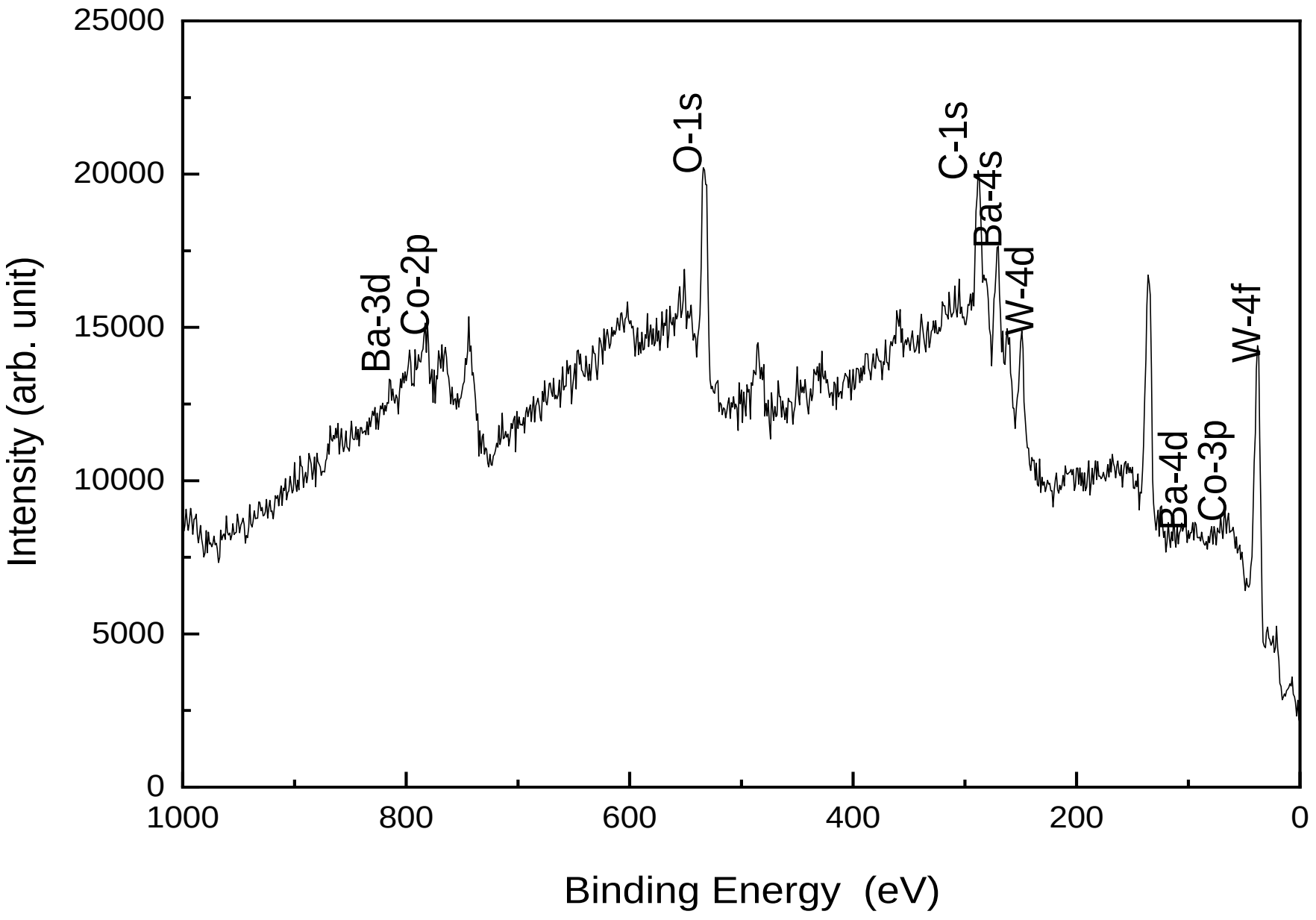

(a)

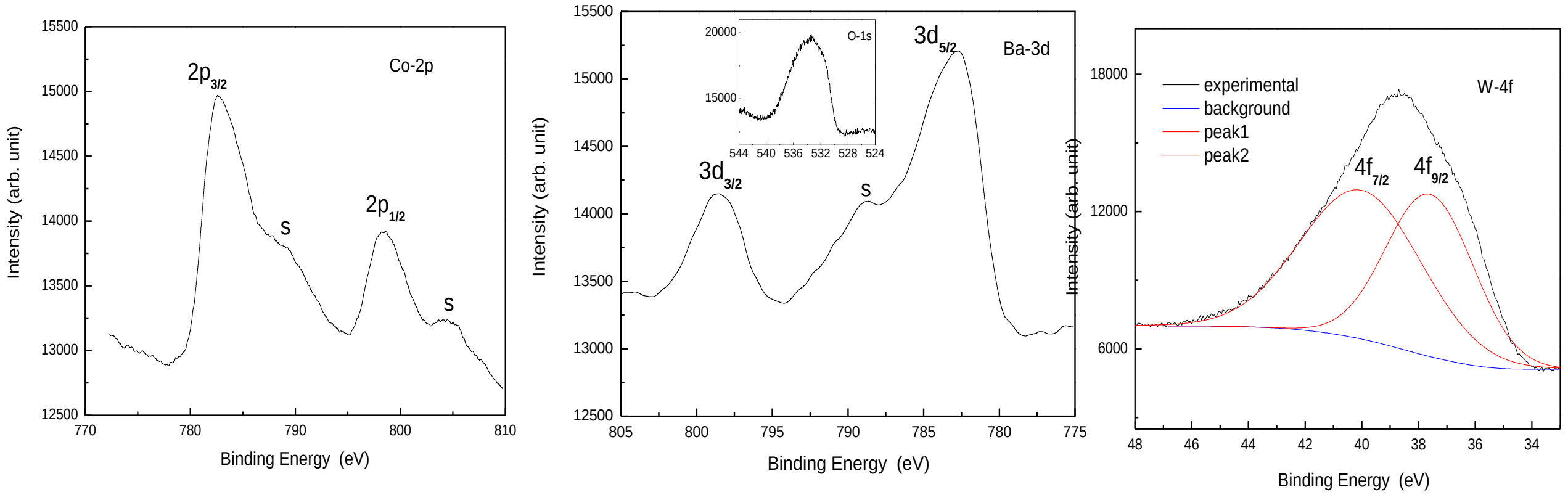

(b) (c) (d)

**Fig. 1:** XPS spectra of BCoW: (a) X-ray photoelectron spectra of BCoW in the energy window of 0 to 900 eV. The core level spectrum of (b) Co 2p, (c) Ba 3d, and (d) W 4f states.

The core level Ba-3d spectrum is shown in Fig. 1(c) The spectrum splits into two components, *viz.* $3d_{5/2}$ and $3d_{3/2}$ because of a spin–orbit interaction. The spin–orbit splitting energy of Ba-3d state is approximately 16 eV with $3d_{5/2}$ (782.6 eV) and $3d_{3/2}$ (798.4 eV) lines at lower and higher binding-energy (BE) positions respectively. The absence of any extra peak in Ba-$3d_{5/2}$ suggests that the existing peak can be assigned for the Ba ion in the perovskite environments. Th presence of a satellite feature (indexed as S) is very common phenomenon in the core level spectrum of alkaline earth metals. The oxygen 1s spectrum in the inset of Fig. 1(c) shows a single peak suggests the presence of only lattice oxygen in the material.

The XPS spectrum of W 4f states is shown in Fig. 1(d) The asymmetric broad nature of the peaks indicates the presence of more than one peak and hence the spectrum is de-convoluted into two peaks. The peak position of $4f_{7/2}$ and $4f_{9/2}$ is 40.0 and 37.6 eV respectively. Since the peaks are at higher binding energy side it indicates the $W^{6+}$ state in this material [17].

### B. Extended X-ray Absorption fine Structure (EXAFS):

The normalized XANES spectra of BCoW and Co foil at Co K-edge have been shown in Fig. 2(a). The spectum for BCoW is very similar to that for CoO [18,19], implying the +2 oxidation state for Co in this compound. The normalized EXAFS absorption spectrum is shown in Fig 1(b). $\chi(R)$ versus $R$ plots at Co K-edge were generated by fourier transform of $k^2\chi(K)$ over $k$ range of 2.5-9.0 $Å^{-1}$ and are shown in Fig. 1(c). The EXAFS data have been fitted by assuming $Ba_2CoWO_6$ crystal structure. The peak at 1.5 Å has contribution of Co-O coordinaton shell while the peak at 3.0 Å has contribution of Co-Ba coordination shell. The best fitted parameters have been listed in Table I.

**Table 1** Bond length, coordination number and disorder factors obtained by EXAFS fitting for $Ba_2CoWO_6$ at Co K-edge.

| **Path** | **Parameter** | **$Ba_2CoWO_6$** |
|---|---|---|
| Co edge | | |
| **Co-O** | R (Å) (2.12) | 2.08±0.005 |
| | N (6) | 6±0.12 |
| | $\sigma^2$ | 0.0048±0.0016 |
| **Co-Ba** | R (Å) (3.51) | 3.43±0.005 |
| | N(8) | 8±0.16 |
| | $\sigma^2$ | 0.0091±0.002 |
| $R_{factor}$ | | 0.005 |

(a
)

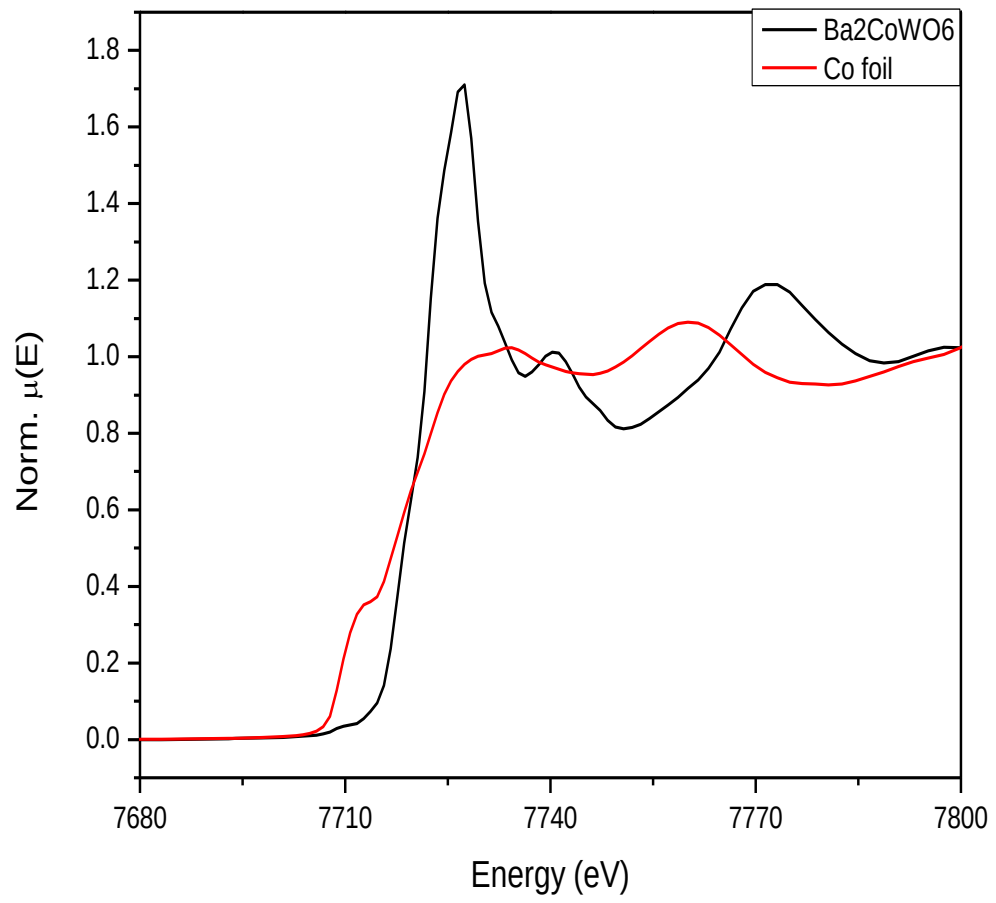

(b
)

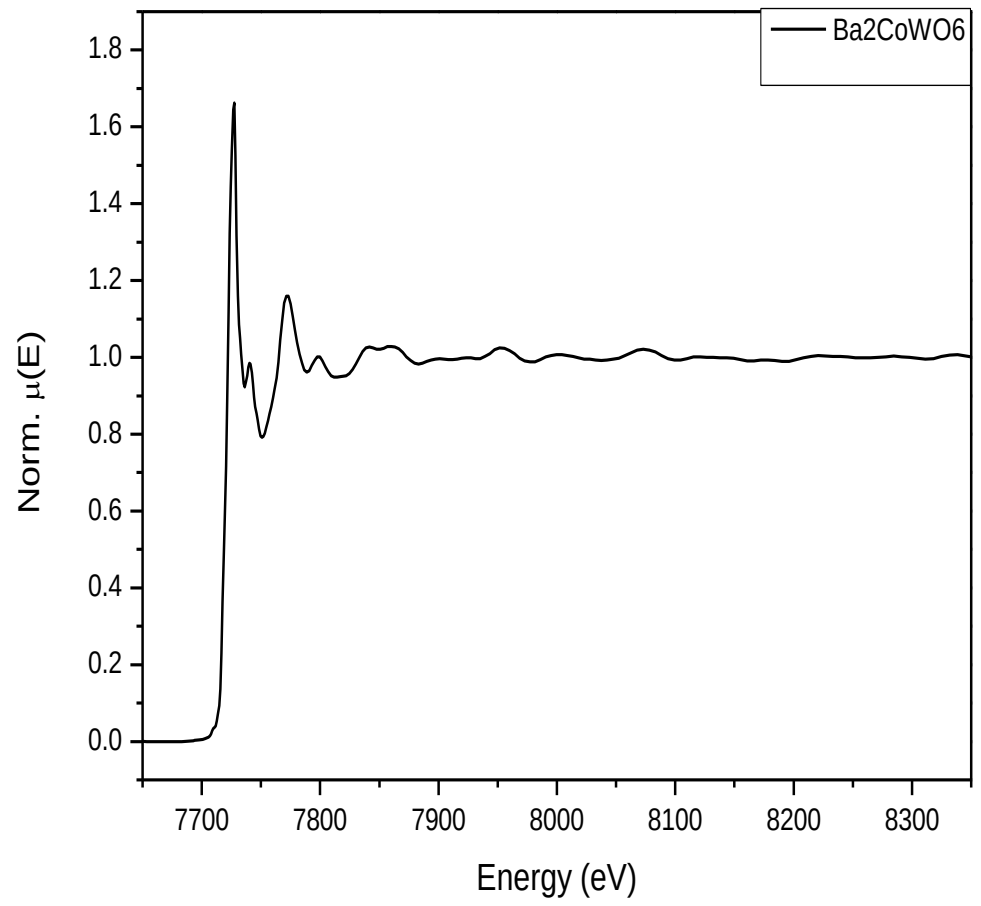

(c)

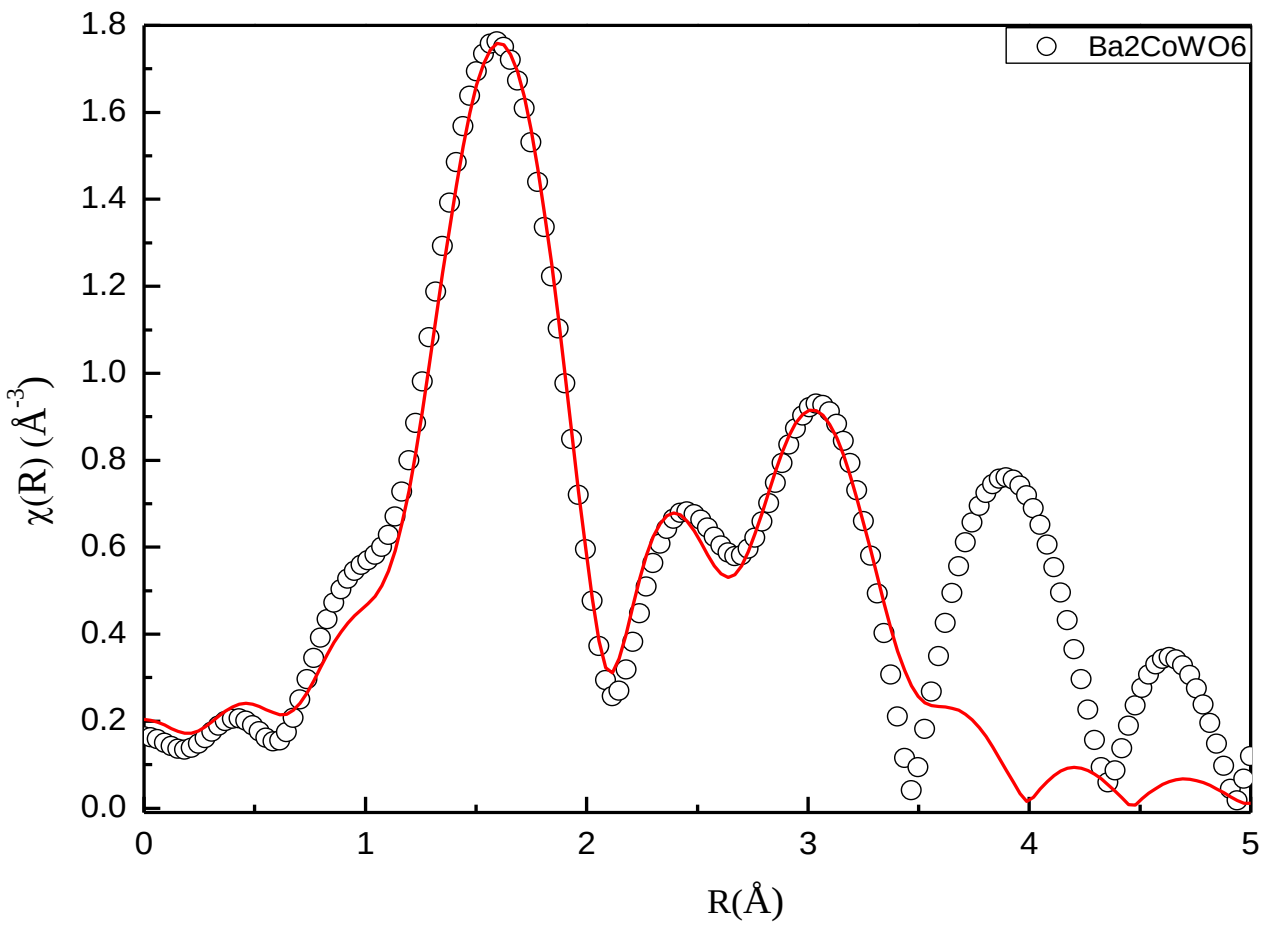

**Fig. 2** (a) Normalized XANES spectra of $Ba_2CoWO_6$ sample and Co foil at Co K-edge. (b) Normalizes EXAFS spectrum of $Ba_2CoWO_6$ sample measured at Co K-edge, and (c) Fourier transformed EXAFS spectrum of $Ba_2CoWO_6$ sample measured at Co K-edge. The experimental spectrum is represented by scatter points and theoretical fit is represented by solid line.